\documentclass[
 reprint,
 amsmath,amssymb,
 prl,
superscriptaddress
]{revtex4-2}

\usepackage{graphicx}
\usepackage[export]{adjustbox}
\usepackage{dcolumn}
\usepackage{bm}

\usepackage{braket}
\usepackage{graphicx}
\usepackage{amsmath}
\usepackage{mathrsfs}
\usepackage{xcolor}
\usepackage{longtable}
\usepackage{amsfonts}
\usepackage{tikz}
\usepackage[colorlinks,linkcolor=blue,citecolor=blue,anchorcolor=blue]{hyperref}
\usepackage{CJK}
\usepackage{upgreek}
\usepackage{tikz}
\usetikzlibrary{quantikz}
\usepackage{xr}
\usepackage{float}

\usepackage{amsmath}
\usepackage[normalem]{ulem}

\begin{document}

\preprint{APS/123-QED}

\title{ 
Fault-tolerant optical  interconnects for neutral-atom arrays
}

\author{Josiah Sinclair}
\affiliation{Department of Physics, MIT-Harvard Center for Ultracold Atoms and Research Laboratory of Electronics, Massachusetts Institute of Technology, Cambridge, Massachusetts 02139, USA}
\author{Joshua Ramette}
\affiliation{Department of Physics, MIT-Harvard Center for Ultracold Atoms and Research Laboratory of Electronics, Massachusetts Institute of Technology, Cambridge, Massachusetts 02139, USA}
\author{Brandon Grinkemeyer}
\affiliation{Department of Physics, Harvard University, Cambridge, Massachusetts 02139, USA}
\author{Dolev Bluvstein}
\affiliation{Department of Physics, Harvard University, Cambridge, Massachusetts 02139, USA}
\author{Mikhail D. Lukin}
\affiliation{Department of Physics, Harvard University, Cambridge, Massachusetts 02139, USA}
\author{Vladan Vuleti\'c}
\affiliation{Department of Physics, MIT-Harvard Center for Ultracold Atoms and Research Laboratory of Electronics, Massachusetts Institute of Technology, Cambridge, Massachusetts 02139, USA}

\date{\today} 

\begin{abstract}

We analyze the use of photonic links to enable  large-scale fault-tolerant connectivity of  locally error-corrected modules based on neutral atom arrays. 
Our approach makes use of recent theoretical results showing the robustness of surface codes to boundary noise and combines recent experimental advances in atom array quantum computing with logical qubits with optical quantum networking techniques.
We find the conditions for fault-tolerance can be achieved with  local two-qubit Rydberg gate and non-local Bell pair errors below 1\%  and 10\%, respectively,  without requiring distillation or space-time overheads.
Realizing the interconnects with a lens, a single optical cavity, or an array of cavities enables a Bell pair generation rate in the 1-50 MHz range.
When directly interfacing logical qubits, this rate translates to error-correction cycles in the 25-2000 kHz range, satisfying all requirements for fault tolerance and in the upper range fast enough for 100 kHz logical clock cycles.

\end{abstract}

\maketitle


\begin{figure*}
\includegraphics[width=\textwidth]{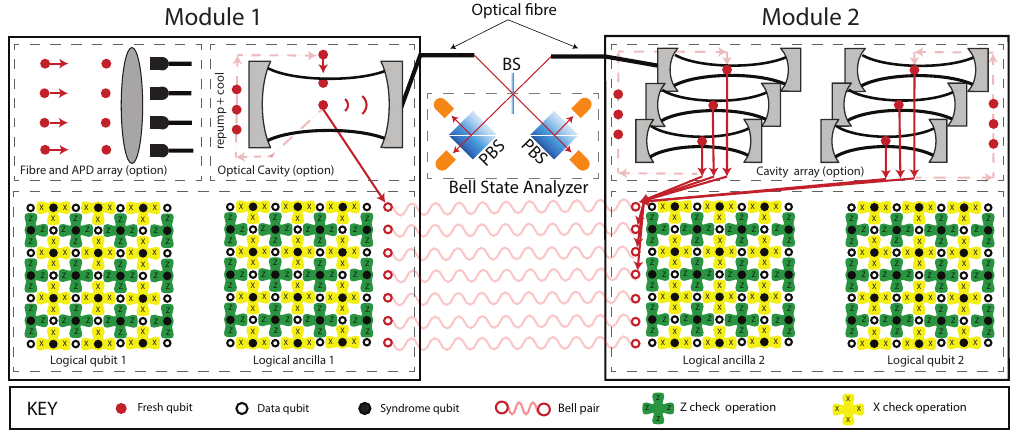}
\caption{Two modules equipped with sufficient quantum input/output for fault-tolerant communication. Surface code patches in each module are realized using an array of atoms and connected using teleported gates.
The Bell pairs are generated using either a lens, an optical cavity, or an array of optical cavities.
Once a Bell pair is created, it is transported (red arrows) from the remote entanglement zone to the computation zone, placed on a logical qubit boundary, and used to enact a parity check between modules.
Interfacing logical qubits in this way allows logical CNOT gates between modules, for example via lattice surgery CNOT gate \cite{horsman2012surface}.
}
\label{rydberg_architecture}
\end{figure*}

Quantum computers made from noisy components require quantum error correction (QEC) to scale \cite{shor1995}.
However, building error-corrected devices is a challenging task, as it requires connecting and controlling a large number of physical qubits with
high fidelity: physical components and operations must introduce noise sufficiently below a code-dependent threshold \cite{Nielsen2011}.
Recently, programmable atom arrays with hundreds of qubits, long coherence times, and high-fidelity Rydberg gates \cite{Madjarov2020, Evered2023} have emerged as a leading platform with experiments demonstrating continuous operation \cite{gyger2024, norcia2024}, control of up to 48 logical qubits, and 
reduction of the error rate with increasing code distance \cite{bluvstein2023}. 
In these experiments, optical multiplexing enables efficient control over logical rather than physical qubits, greatly easing experimental complexity and paving the way for intermediate-scale error-corrected devices containing hundreds of logical qubits.

Further scaling will greatly benefit from modular architectures, where modules containing many physical qubits and supporting high-fidelity and error-corrected `local' quantum operations are connected via long-range entanglement, mediated by photons \cite{Nickerson2013, Monroe2014, Nickerson2014}.  
Modular architectures have the appeal of transforming the challenge of scaling into the task of realizing a unit module with a sufficiently high-rate and fidelity quantum input-output interface for fault-tolerant communication. 
Once this is accomplished, the system can then be scaled up arbitrarily by adding more modules, eliminating the need to solve new scientific and engineering challenges at each new larger size.

In this paper, we propose and analyze such a unit module for fault-tolerantly linking Rydberg arrays via photonic interconnects. 
We first consider the noise requirements for fault-tolerant quantum communication.
While multiple recent works have targeted network noise below $1\%$ to achieve modular fault-tolerance \cite{li2024highrate, leone2023, sunami2024, singh2024}, we show the network noise can in fact exceed 10\%.
We apply to neutral atoms the recent results of \cite{ramette2023}, which showed that distinct surface code patches can be connected in a fault-tolerant manner even in the presence of substantial noise along their connecting interface. 
Translated to a Rydberg qubit error model, we find this corresponds to a local Rydberg gate and non-local Bell pair (photonic interconnect) threshold of $1\%$ and $10\%$, respectively, representing a tenfold increase in the Bell pair threshold and comparing favorably to the current state of the art 
\cite{Evered2023, Wilk2010, Isenhower2010, Jau2016, Graham2019, Levine2019, Bluvstein2022, Madjarov2020, Maunz2009, Stephenson2020, Krutyanskiy2023, Nolleke2013, Langenfeld2021}.
This high tolerance for communication errors brings the fault-tolerant connection of error-corrected modules within reach of existing atom array technology, with the primary remaining challenge being the development of sufficiently fast and efficient photonic links. 

Turning to this challenge, we next explore several module designs depicted in Fig. \ref{rydberg_architecture} that aim to achieve fast quantum communication. 
Our module designs utilize a zoned architecture \cite{bluvstein2023}, leveraging the recently demonstrated capability of fast coherent transport \cite{Lengwenus2010, Schlosser2011, Bluvstein2022}.
Bell pairs are generated in a ``remote entanglement generation zone" and then transported to a quiescent ``computation zone" where logical qubits are maintained and operated. 
In the first design, a large numerical aperture (NA) lens and detector array are used to entangle atoms remotely.
In this approach, the rate of entanglement generation, $\sim 200$ Hz/atom, is limited by the lens's collection efficiency, but very high rates can be reached with sufficient multiplexing.
Contrasting with this slow but highly parallel approach, we consider a design employing a single optical cavity using the realistic parameters from Young et al. \cite{Saffman2022}. 
Parallel control of multiple itinerant qubits combined with temporal multiplexing can enable Bell-pair generation rates reaching $\sim 1$ MHz and error correction cycles of $25$ kHz.
Finally, we seek to combine the parallelism of the free-space approach with the speed of an optical cavity by introducing a third module design that utilizes an array of optical micro-cavities. 
For an array of 30 cavities, we estimate a maximum Bell pair generation rate of approximately $50$ MHz, capable of $2$ MHz error correction cycles. 
Estimating the number of qubits needed for high-speed fault-tolerant communication, we discuss trade-offs between communication speed and qubit overheads in each approach.

\section{Communication fidelity requirements for fault-tolerant scalability}

A significant challenge for all quantum error-corrected modular architectures is noise in the quantum communication between modules.
Typically, architectures based on the surface code target both local gate and network noise levels of 0.1\% \cite{li2024highrate, leone2023, sunami2024, singh2024} to ensure all operations occur around 10$\times$ below the usual surface code threshold of 1\%.
While Rydberg gates have recently surpassed the $1\%$ threshold, the lowest Bell pair infidelity achieved to date in atomic qubits lags behind at $6-10\%$  \cite{Stephenson2020, Hartung2024} (see \cite{covey2023quantum} for a review). 
Additionally, teleported gates require several local operations to be executed on the Bell pair, further increasing network noise.
To reduce network noise, entanglement distillation has been proposed as a method to convert many low-fidelity Bell pairs into a smaller number of higher-fidelity Bell pairs \cite{Dur2003}.
However, even simple protocols involve many local gates and, therefore, can only distill down to an error rate approximately ten times larger than the local gate errors \cite{Campbell2007, Krastanov2019}.
More sophisticated protocols are necessary to reduce the noise further, but the added complexity lowers the success rate,
reducing the achievable error correction cycle rate and increasing the memory errors per cycle
\cite{Liang2007, Li2012, Fujii2012, Nickerson2013, Nickerson2014, Nigmatullin2016}.

\begin{figure*}
\includegraphics[width=18cm]{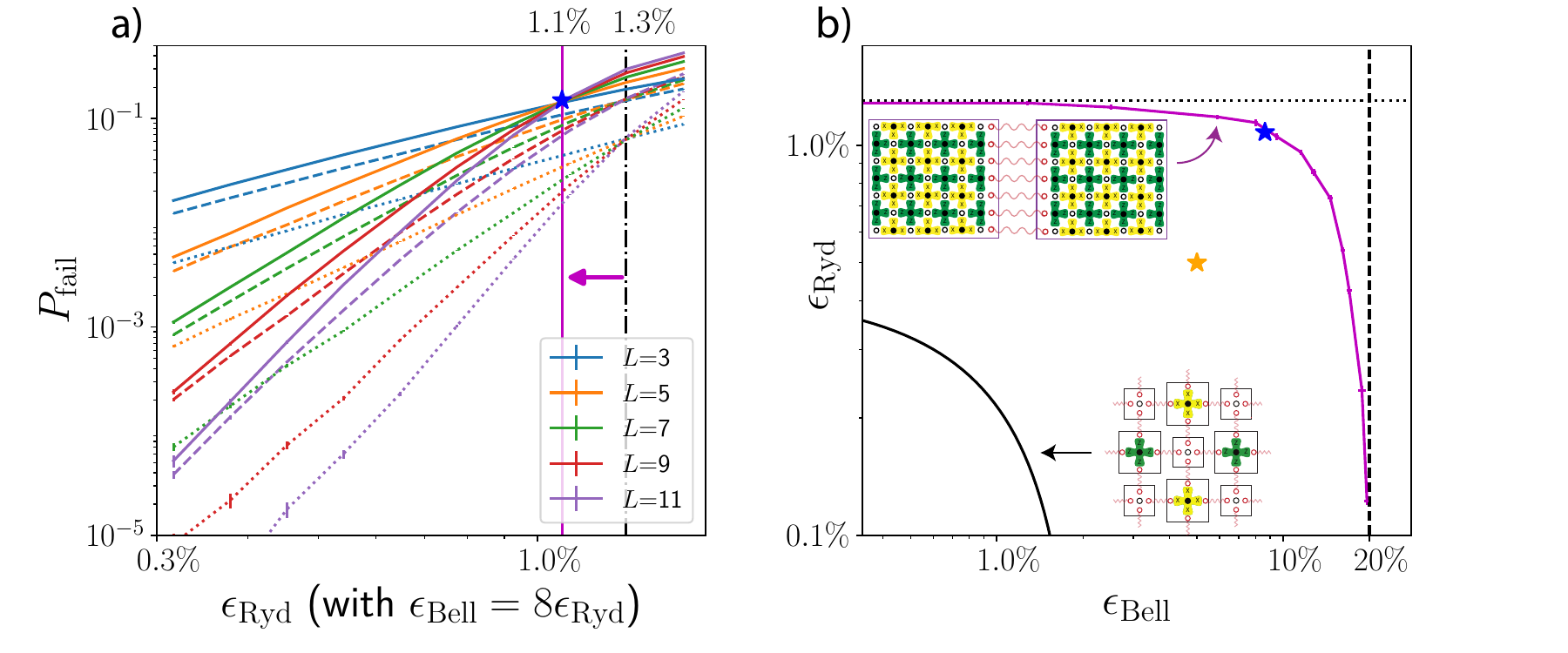}
\caption{
(a) Dotted: $P_\textrm{fail}$ for a bulk surface code patch with \textbf{no boundary} errors (in terms of $\epsilon_\textrm{Ryd}$). 
Dashed: plotting \textbf{only the boundary} logical errors in terms of $\epsilon_\textrm{Ryd}$, fixing $\epsilon_\textrm{Bell} = 8 \epsilon_\textrm{Ryd}$, $\epsilon_\textrm{M} = \epsilon_{\textrm{Ryd}}$ to shift the boundary threshold to equal the bulk threshold (black dotted-dashed vertical line).
Solid: $P_\textrm{fail}$ with \textbf{both bulk and boundary} errors simultaneously. 
(b) Magenta: the threshold for $P_\textrm{fail}$ plotted in the $(\epsilon_\textrm{Ryd}, \epsilon_\textrm{Bell})$ space.
Points below and left of the curves are below threshold.
Blue star indicates threshold in (a), orange star indicates best experimental Rydberg gate and non-local Bell pair infidelity achieved separately to date \cite{Evered2023, Stephenson2020}.
Black: for reference, the curve 
$3\% = 1.5 \epsilon_{\textrm{Bell}} + 7 \epsilon_{\textrm{Ryd}}$, 
approximates the expected threshold if all parity checks in the bulk were carried out with teleported gates.
}
\label{threshold_shifts}
\end{figure*}

As an alternative to distillation, it has been previously shown \cite{ramette2023, fowler2010, tan2024} that the surface code is intrinsically robust to higher levels of noise along a lower-dimensional boundary.
As shown in Fig. \ref{rydberg_architecture}, this boundary is where the physical operations required to fault-tolerantly connect code patches occur, for example, using lattice surgery to perform a logical CNOT gate \cite{horsman2012surface}.
For noise confined to the boundary, the boundary can be thought of as acting like an embedded lower-dimensional code with correspondingly less stringent thresholds \cite{Dennis2002, fowler2010, ramette2023, tan2024}.
Regardless of the degree of noise on the boundary and the bulk, most logical failures are isolated to either the lower dimensional boundary or the bulk.
Consequently, the thresholds for boundary and bulk gates are almost entirely independent.
One can then approximate the total logical error rate as arising from logical errors occurring on the bulk and on the boundary independently:
\begin{align}\label{simplemodel}
    P_\textrm{fail}(p_\textrm{bulk},p_\textrm{bound}) \sim \Big( \frac{p_\textrm{bulk}}{p_\textrm{bulk}^\textrm{th}} \Big)^{L/2} + \Big( \frac{p_\textrm{bound}}{p_\textrm{bound}^\textrm{th}} \Big)^{L/2},
\end{align}
where errors not confined to either the bulk or the boundary are negligible, and the thresholds obey $p_\textrm{bound}^\textrm{th} > p_\textrm{bulk}^\textrm{th}$. 
From Eq. \ref{simplemodel}, we can see that errors on the boundary will not dominate the total error until $p_\textrm{bound}/p_\textrm{bound}^\textrm{th}$ is comparable to $p_\textrm{bulk}/p_\textrm{bulk}^\textrm{th}$.

Given the long coherence time of neutral atoms, errors occur primarily due to imperfect local Rydberg gates (with probability $\epsilon_\textrm{Ryd}$), local fluorescence readout ($\epsilon_\textrm{M}$), and non-local Bell pairs ($\epsilon_\textrm{Bell}$).
We model this noise by counting up the probabilities $p$ and $q$ per error correction cycle for the sum of these processes to create a bit flip error on data and syndrome qubits respectively, both in the bulk and on the boundary (see appended supplement for more information about the error model), with the results summarized in Table \ref{table1}. While noise in the bulk comes solely from local operations necessary to measure the local check operators, Bell pair noise and additional local operations for enacting the nonlocal teleported gate circuit contribute to elevated noise along the boundary.

\begin{table}[h!]
\centering
\begin{tabular}{|c | c |c|} 
\hline
 & Bulk & Boundary \\ 
 \hline
 $p$ & $2 \epsilon_\textrm{Ryd}$ & $0.5 \epsilon_\textrm{Bell}+2.5 \epsilon_\textrm{Ryd}+ \epsilon_\textrm{M}$ \\ 
 \hline
 $q$ & $2 \epsilon_\textrm{Ryd}+ \epsilon_\textrm{M}$ & $0.5 \epsilon_\textrm{Bell} +2.5 \epsilon_\textrm{Ryd}+2 \epsilon_\textrm{M}$ \\
\hline
\end{tabular}
\caption{Bit flip error rates associated with bulk and boundary qubits under a generic, unbiased noise model.}
\label{table1}
\end{table}

Using this Rydberg error model, we perform Monte Carlo simulations of surface codes with phenomenological errors in the bulk and on the boundary \cite{Higgott2021}, and plot the logical failure rate $P_\textrm{fail}$ in Fig.~\ref{threshold_shifts}a in terms of $\epsilon_\textrm{Ryd}$, choosing $\epsilon_\textrm{Ryd} = \epsilon_\textrm{M}$ and having fixed the ratio $\epsilon_\textrm{Bell} = 8 \epsilon_\textrm{Ryd}$.
The cases of only boundary noise (dashed), only bulk noise (dotted), and both boundary and bulk noise occurring simultaneously (solid) are each shown.
The elevated noise on the boundary, due to both the additional local operations necessary to operate the teleported gate as well as the noise from the Bell pairs, only modestly shifts the $\epsilon_\textrm{Ryd}$ threshold. 
Fig.~\ref{threshold_shifts}b shows how the same threshold depends on the choice of parameters $\epsilon_\mathrm{Ryd}$, $\epsilon_\mathrm{Bell}$, with the blue star identifying the point $\epsilon_\mathrm{Ryd} = 1.1 \%, \epsilon_\mathrm{Bell} = 8.8 \%$ corresponding to the threshold in Fig. \ref{threshold_shifts}a. 
To highlight how favorable these threshold requirements are, the orange star identifies the best remote atomic Bell pair and Rydberg gate fidelities experimentally achieved so far in separate experiments \cite{Stephenson2020, Evered2023}, while, for comparison, the black line shows the expected thresholds if each module contains only a couple of qubits and all parity checks in the bulk are carried out with teleported gates (see appended supplement for a detailed description of the error model).

\section{Communication speed requirements for fault-tolerant scalability}

Having shown a high degree of robustness to noise along the interface between modules, successful error correction now only requires executing error correction cycles sufficiently fast.
To ensure that background decoherence errors remain negligible compared to other noise sources, we require a cycle time $T$ satisfying $T/\tau_{\textrm{dec}} \ll p_\textrm{bulk}$, where a decoherence time $\tau_{\textrm{dec}} \sim 2$ second has been demonstrated \cite{Ma2021, Bluvstein2022}.
Beyond decoherence, faster error correction cycles are desirable for improving logical clock speeds. 
The rate of generating the $2L$ Bell pairs required per cycle, where $L$ is the surface code distance, is expected to bottleneck $T$. 
(The fastest experimentally achieved remote entanglement generation rate in neutral atoms or trapped ions is currently $250$ Hz \cite{oreilly2024}.)
However, through either multiplexing entanglement generation in free space, or through the use of a single optical cavity we show that it should be possible to generate the $2L$ Bell pairs per error correction cycle much faster than necessary to interface logical qubits of size and quality sufficient for large-scale applications.
We also show that even higher rates are possible if entanglement generation can be further parallelized using an array of micro-cavities and compare rates and qubit overhead for each approach.

For generating non-local Bell pairs, we consider remote entanglement protocols where atom-photon entanglement is swapped to atom-atom entanglement \cite{Pan1998, Schupp2021, Stephenson2020, oreilly2024}.
Atom-atom entanglement is probabilistic but heralded by the successful detection of two photons.
For concreteness, we consider the Rubidium atom-atom entanglement protocol from \cite{Saffman2022,hofman2012, rosenfeld2017},  operating on the D$2$ $F = 1$ to $F' = 0$ transition.

\subsection{Free-space collection}
Considering first just a single atom, the single-photon emission success probability is given by the mode overlap of the dipole electric field of the atom and a Gaussian mode. 
Assuming a numerical aperture of $0.7$, the collection efficiency $\eta ^\textrm{lens} = 0.12$ \cite{Saffman2022}.
The atom-atom entanglement generation probability per attempt $P_\textrm{aa}$ is then given by
\begin{align}
    P_\textrm{aa}^{\textrm{lens}} = \frac{1}{2} \big(\eta^\textrm{lens} \eta_\textrm{det} \big)^2,
\end{align}
where $\eta_\textrm{det}$ is the subsequent detection efficiency.  
We assume $\eta_\textrm{det} = 0.7$, making the atom-atom entanglement generation success probability per attempt $P_\textrm{aa}^\textrm{lens} = 0.0035$.

With a single qubit, the repetition rate is limited by the time required to reset the qubit after each failed attempt (about $6 \: \upmu$s) and the time required to occasionally recool the qubits (approximately $10 \: \upmu$s per attempt) \cite{Saffman2022}. 
The resulting mean time to successfully prepare a Bell pair is then $16 \: \upmu$s$/0.0035 = 4.6$ ms.
Faster rates are possible using a workflow where atom re-initialization and cooling are carried out elsewhere to allow entanglement attempts to be repeated more frequently. 
In addition, entanglement can be attempted on multiple atoms simultaneously using a detector array (see Fig. \ref{speed_compare}). 
In this case, the Bell pair generation rate is limited only by the number of atoms which are available and which can be simultaneously controlled.

As summarized in Table \ref{table3}, the cycle time $T$, dominated by the time to distribute Bell pairs, will then be $T = 2L \tau_{\textrm{Bell}}$, the time to create the $2L$ Bell pairs along the interface per cycle.
$T$ grows with $L$, but as a reference point, with local gates and boundary noise operating $10 \times$ below the threshold, as in Eq. \ref{simplemodel}, logical qubits with distance $L=20$ have logical errors sufficiently low for large-scale applications \cite{Fowler2012}. 
In this case, $\tau_{\textrm{Bell}} = 4.6$ ms/atom, so control of $\sim 100 $ qubits is sufficient to produce $40$ Bell pairs in $2$ ms, fast enough so that $T/\tau_{\textrm{dec}} = 10^{-3}$. Implementing a logical gate requires $L$ code cycles, giving a logical clock speed of $25$ Hz.

\begin{table}[h!]
\centering
\begin{tabular}{|c | c | c | c | c |} 
 \hline 
 & & Free-space & 1 cavity & 30 cavities \\
 \hline
 Code distance & $L$ & 20 & 20 & 20 \\ 
 \hline
 Qubit \# $N$ &$\sim 4 L^2$ & 1521 & 1521 & 1521\\ 
 \hline
 \# of BP's  &$2 L^2$ & 800 & 800 & 800 \\
 \hline
 BP attempt time & $\tau$ & $0.10 \: \upmu$s/qubit & $0.10 \: \upmu$s  & $3.3$ ns\\ 
 \hline
 BP time $\tau_{\textrm{Bell}}$ & $\tau/P_{\textrm{aa}}^{\textrm{cav}}$ & $4.6$ ms/qubit & 1.0 $\upmu$s &  $14$ ns\\ 
 \hline
 Cycle time $T$ & $2L \tau_{\textrm{Bell}}$ & 180 ms/qubit  & 40 $\upmu$s & $560$ ns \\
\hline
Log. gate time & $ 2 L^2 \tau_{\textrm{Bell}}$ & 3.7 s/qubit & 0.8 ms & $11.2 \: \upmu$s \\
\hline
 Decoh. time & $\tau_{\textrm{dec}}$ & 2.0 s & 2.0 s & 2.0 s\\
\hline
\end{tabular}
\caption{Quantities relevant for estimating QEC cycle rates for a code patch with distance $L=20$ when using a lens, a single cavity, or a micro-cavity array with the parameters from Table \ref{table2}. For free space, times are reported per communication qubit. Where an optical cavity is used, the times reported are the optimal value from Fig. \ref{speed_compare} limited by the number (1 or 30) of optical modes. Bell pairs are denoted BP.} 
\label{table3}
\end{table}


\subsection{Single large-volume cavity}
By enhancing collection efficiency, an optical cavity can create Bell pairs faster than in free space, enabling faster logical clock speeds with fewer communication qubits.
We adopt the ``medium near-concentric" cavity design proposed in \cite{Saffman2022}, where to be compatible with Rydberg excitation the cavity length is conservatively chosen to be $4$ mm with a $5 \: \upmu$m waist. 
The remaining experimental parameters are summarized in Table \ref{table2}. 
The cavity enhances collection efficiency from the atom, achieving $\eta^{\textrm{cav}} = 0.66$.
Assuming the same detection efficiency of $0.7$, the atom-atom entanglement generation probability per attempt is $P_\textrm{aa}^{\textrm{cav}} = 0.1$. 

\begin{table}
\centering
\begin{tabular}{|c | c | c |} 
  \hline
   & single cavity & micro-cavity array \\
   \hline
   \# of cavities & 1 & 30 \\
  \hline
 Length & 4 mm  & 90 $\upmu$m\\ 
  \hline
 Waist & 5 $\upmu$m  & 2.5 $\upmu$m\\ 
 \hline
 (g, $\kappa$, $\Gamma$) & ($17, 28, 6$ MHz) & ($184, 404, 6$ MHz) \\
 \hline
 Cooperativity $C$ & 3.2 & 56\\ 
 \hline
 $\eta ^\textrm{cav}$& 0.66 & 0.98\\ 
 \hline
 $\eta_\textrm{det}$ & 0.7 & 0.7\\
\hline
 $P_\textrm{aa}^\textrm{cav}$ & 0.1 & 0.24\\ 
 \hline
$\tau_\textrm{cav}$ & 5.7 ns & 0.39 ns\\
\hline
\end{tabular}
\caption{Cavity properties: for the single cavity, we adopt the ``medium near-concentric" entry from Table II of \cite{Saffman2022}. For the micro-cavity array, we envision a cavity with a radius-of-curvature of $\sim 100 \: \upmu$m, allowing for a large cavity bandwidth and small mode volume. The small size of the mode allows for these cavities to be densely packed.}
\label{table2}
\end{table}



If only one qubit is available for entanglement generation, Ref. \cite{Saffman2022} previously found that entanglement rates of about $5800$ Hz are possible, limited by the $16 \: \upmu$s required to reset and occasionally recool qubits between subsequent attempts.
To achieve faster rates of remote entanglement, we envision a module design that utilizes the coherent transport capabilities of neutral atoms \cite{Lengwenus2010, Schlosser2011, Bluvstein2022} to temporally multiplex remote entanglement generation \cite{krutyanskiy2023multimode, sunami2024}. 
In each module, depicted in Fig. \ref{rydberg_architecture}, fresh qubits are initialized and transported through the cavity, where attempts are made to establish remote entanglement.
If successful, the qubits comprising the Bell pair are transported to the edge of the code patch (red lines in Fig. \ref{rydberg_architecture}) where they will be used to enact syndrome checks across the boundary.

The speed at which atoms can be transported through the cavity mode is limited by trap depth (available laser power) and heating to approximately $1$-$10$ $\upmu$m/$\upmu$s \cite{Bluvstein2022}. 
With a 5 $\upmu$m waist, even at top speed atoms will spend much longer in the cavity mode than is required to attempt remote entanglement.
To increase the attempt rate, off-resonant addressing beams focused to 1 $\upmu$m can be used to Stark shift atoms as they exit and enter the central region of the cavity mode, allowing for a more closely-spaced stream of qubits and an attempt every $100$ ns.

\begin{figure}[t!]
\includegraphics[width=8.6cm]{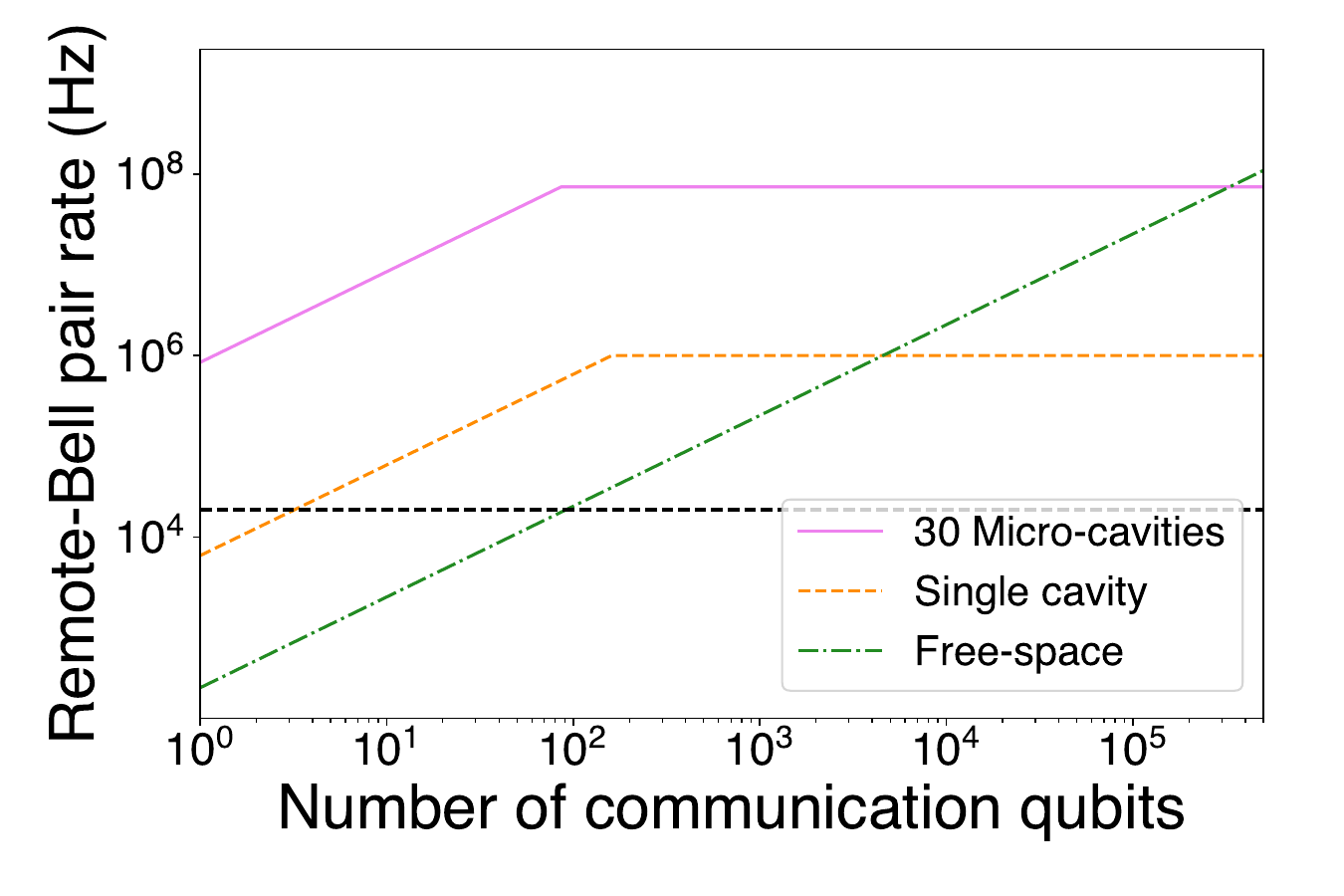}
\caption{The average Bell pair rate depends on the number of qubits available for communication and the interconnect design. 
The dashed black line indicates the rate required to execute error correction cycles every 2 ms for one distance $L=20$ logical qubit.
For a single cavity (orange dashed line), above $160$ qubits the rate becomes limited by addressing beam switching time. Higher rates (purple solid line) are possible using an array of micro-cavities. In this case, the rate is eventually limited by the switching time and by the number of micro-cavities, which larger microscope objectives could further increase.
In free space (green dot-dashed line), lower collection efficiency means more qubits are necessary to achieve the same Bell pair rate.}
\label{speed_compare}
\end{figure}

An additional possibility proposed in \cite{li2024highrate} is to transport qubits in batches into the cavity and attempt remote entanglement sequentially (relying again on individually addressed Stark shifts). 
Applying these methods with about $160$ communication qubits (see Fig. \ref{speed_compare}) should make it possible to reach an attempt rate limited by the switching time of the addressing beams, enabling remote entanglement attempts approximately every $100$ ns.
With a $P_{aa}^{\textrm{cav}}$ success probability of $0.1$, the maximum mean Bell pair time limited by there being only one optical mode is $\tau_{\textrm{Bell}} = 1 \: \upmu$s. 
The time to produce the $40$ Bell pairs required to implement an error correction cycle is then $T = 40 \: \upmu$s, resulting in a $25$ kHz code cycle rate and a logical clock speed of $1.25$ kHz.

\subsection{Micro-cavity array}
With one cavity, the use of a single optical mode presents a fundamental limit on the Bell pair rate, as entanglement attempts must be made sequentially.
This limit can be increased through interfacing multiple cavities to the same array \cite{Trupke2007, Derntl2014, Wachter2019}.
However, field-of-view constraints require a different approach to cavity design. 
The small size of micro-cavities allows for them to be densely packed into a micro-cavity array. 
The main considerations for the density are the clipping loss of the cavity waist on the micro-cavity mirror and the total length of the cavity including the substrate.

We propose micro-cavities requiring a $50 \: \upmu$m mirror diameter and a substrate less than $100 \: \upmu$m thick. With a $1.5$ mm field-of-view microscope objective, about 30 cavities can be fit side-by-side. Stabilization can then be performed through a combination of micro-electromechanical systems (MEMS) and thermal tuning to allow for independent locking of each high finesse cavity \cite{derntl2014arrays, gallego2016high}.
The resulting cavities are $90 \: \upmu$m in length, and are tightly focused with a waist of $2.5 \: \upmu$m enabling a large cooperativity of $C=56$. 
To ensure compatibility with Rydberg excitation, doped silicon should be used for the substrate, reducing charge noise by making the bulk of the structure conductive. 
Additionally, if necessary, Rydberg gates can be performed in the computation zone away from the cavities.
For a summary of their other properties, see Table \ref{table2}. 

Because the time to attempt entanglement and extract a photon from the cavity is shorter than all other timescales ($1 / \kappa \sim 1$ ns), it would be beneficial to attempt remote entanglement on each qubit multiple times while it is in the cavity mode. 
This is impractical with free-space optical pumping of the atom, which requires approximately $6 \: \upmu$s. It can be achieved, however, by
making use of Purcell-enhanced decay to perform fast state preparation.
In the case of entanglement generation with a single atom, this gives an effective increase in entanglement generation rates by roughly a factor of the cooperativity.

For rubidium, we propose a cavity-enhanced pumping scheme which can reduce the optical pumping time from $6 \: \upmu$s to $100$ ns with $>$ 99$\%$ fidelity (see appended supplement for more details). 
This pumping is performed on the D1 $F=1 \leftrightarrow F'=1$ transition.
Selection rules forbid the $m_F = 0 \leftrightarrow m_F'=0$ transition, resulting in the accumulation of population into the $|F=1, m_F=0\rangle$ state. 
With a cavity resonance on the $F=1 \leftrightarrow F'=1$ D1 transition, the atom will have an enhanced scattering rate from the $|F'=1,m_F'=\pm 1\rangle$ states into the $|F=1,m_F = 0\rangle$. 
This enhanced scattering rate pumps the atom more rapidly while also reducing the total number of scattered photons required to pump the atom, reducing the amount of required cooling by a similar factor of $C$.
By reducing the time required to repump and recool qubits, Purcell-enhanced pumping reduces the number of qubits needed to reach a Bell pair attempt rate limited by the number of optical modes by $C$.
For comparison, $160$ communication qubits are needed to achieve a Bell pair attempt every $100$ ns with a single cavity, but only $\sim 3$ are needed to achieve a similar rate in a single micro-cavity with cooperativity $C = 56$. 
With an attempt every $100$ ns on each cavity, $30$ cavities,  and $100$ communication qubits, an attempt can be made on average every $3.3$ ns. 
Per  Table \ref{table3}, the average time to create a Bell pair is then $\tau_{\textrm{Bell}} = 14$ ns, resulting in a $2$ MHz QEC cycle rate and a logical clock speed of $100$ kHz.

\subsection{Comparison of rates and qubit resources}

In Fig. \ref{speed_compare}, we compare the qubit resources required to achieve a certain remote entanglement rate for our three approaches.
We plot the remote Bell pair rate versus the number of communication qubits for a lens (dot-dashed line), a single large-volume cavity (dashed line), and a micro-cavity array (solid line).
For the comparison, the cavity parameters used are given in Table \ref{table2} (see appended supplement for a detailed description of the rates shown).
The corners at around $160$ communication qubits for the single large-volume cavity and around $100$ communication qubits for the micro-cavity array show the point at which the rate becomes limited by the number of optical modes.
The dashed black line indicates a Bell pair rate sufficiently fast to execute error correction cycles every 2 ms for one distance $L=20$ logical qubit, satisfying the requirement that $T/\tau_{\textrm{dec}} = 10^{-3}$.

\section{Supermodules}

To reduce the number of total modules required, super-modules can be constructed comprising multiple arrays in one vacuum chamber connected by optical lattice conveyor belts. 
Each array (or ``sub-module") would be separately controlled and under its own microscope, with the crucial challenge being fast and high-fidelity communication between sub-modules via atom transport. 
For the former, many Bell pairs would be created deterministically using parallelized local Rydberg gates. Then, one qubit out of each pair would be accelerated to match the speed of the optical lattice conveyor belt, placed in the lattice, and transported to a distant array. 

Because Bell pairs can be generated using Rydberg gates, they could be initially prepared with fidelity exceeding 99\% \cite{Evered2023}.
Therefore, the predominant source of Bell pair infidelity would be decoherence during transport. 
Transport times are set by the conveyor belt speed, which is limited by the maximum velocity that qubits can reach without excessive heating ($\sim 1 \: \upmu$m/$\upmu$s). 
Travel times between arrays separated by 10 cm (limited by space required for microscope objectives) would therefore be approximately 100 ms, resulting in decoherence below the $10\%$ Bell pair threshold from Fig. \ref{threshold_shifts}b.

\section{Conclusion}

Combining unique atom array capabilities with the recently reported robustness of surface codes to boundary noise, we have proposed a path to fault-tolerant scaling of error-corrected modules requiring only current-to-near-term technology.  
Rydberg gate errors below $1\%$ have been achieved  \cite{Evered2023} as well as remote entanglement of atomic qubits with infidelity below $10\%$ \cite{Stephenson2020, oreilly2024}. 
Additionally, theoretical analysis of the expected fidelity for remote Bell pairs indicates further improvements can be expected \cite{Saffman2022, li2024highrate}.
Generating Bell pairs sufficiently fast remains a challenge \cite{Saffman2022, covey2023quantum, li2024highrate}, but we have described several promising approaches for realizing sufficient Bell pair rates using free-space collection, a large-volume optical cavity, or a micro-cavity array. 
Injecting physical Bell pairs into logical qubits \cite{sunami2024} and running more sophisticated distillation schemes \cite{gidney2023} could require fewer physical Bell pairs per logical gate, but would incur additional space-time overheads, and further work is warranted to investigate these tradeoffs.
Further gains may also be possible in certain regimes by using transversal gates between modules  \footnote{See Supplementary Fig. 3 of [1]} and exploiting algorithmic fault tolerance \cite{cain2024, zhou2024}.
By lowering the bar for communication fidelity and outlining multiple paths towards achieving sufficiently fast photonic interconnects, the present analysis motivates the realization and exploration of networked logical quantum processors that satisfy the requirements for scalable fault tolerance in the near term.

\section*{Acknowledgements}
This project was funded in part by DARPA under the ONISQ program (grant \# W911NF2010021), the US Department of Energy (Quantum Systems Accelerator, grant \# DE-AC02-05CH11231), the Center for Ultracold Atoms (an NSF Physics Frontiers Center, Grant \# PHY-1734011), QuSeC-TAQS from NSF (grant \# 2326787), 
(grant \# W911NF2320219), the Army Research Office (grant \# W911NF2010082), IARPA under the ELQ program (grant \# W911NF2320219), and QuEra Computing. 
D.B. acknowledges support from the NSF Graduate Research Fellowship Program (grant \# DGE1745303) and The Fannie and John Hertz Foundation.

\bibliography{MAIN}

\section*{Supplemental Materials}

\subsection{Teleported gate}
\begin{figure}[h]
\includegraphics[width=0.5\textwidth]{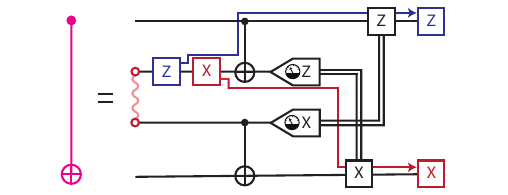}
\caption{X, Z errors on the Bell pair used in teleported gate propagate to the two qubits it operates on.}
\label{circuit}
\end{figure}

When the interface is realized via distributed entanglement, that entanglement serves as a resource for enacting non-local, teleported gates between qubits in distinct modules.
Fig.~\ref{circuit} shows how bit and phase flip noise on the distributed Bell pair propagates to the control and target qubits in the distinct modules that the (pink) teleported gate acts on.
The propagation is identical for errors occurring on either of the Bell pair qubits, as must be the case since the Bell pair is invariant under application of $XX$ and $ZZ$.

\subsection{Microscopic Seam Error Model}

For Bell pairs and CNOT gates, we model each as a perfect operation followed by all combinations of $X, Y, Z$ errors on the two qubits, each with error $\epsilon_\textrm{B}/15$ or $\epsilon_\textrm{CX}/15$: $IX, IY, IZ, XI, XX, XY,$ $XZ, YI, YX, YY, YZ, ZI, ZX, ZY, ZZ$ \cite{fowler2010}. As the Bell pair is stabilized under the application of $XX, ZZ$, all 15 types of errors are equivalent to $II$ with $\frac{1}{5} \epsilon_\textrm{B}$ and $IX$, $IY$, or $IZ$ with probability $\frac{4}{15}\epsilon_\textrm{B}$.
Similarly, as $Y \propto XZ$, we can count up the total probability a given qubit gets hit by $X$ and by $Z$ errors following a CNOT: $\frac{8}{15} \epsilon_\textrm{CX}$.

Using this error model, we first consider a standard surface code patch. In this case there is no seam and errors only arise from CNOT gate errors and readout errors. As each data qubit is subject to four parity check operations per code cycle, the bit flip probability for data qubits is  $p_\textrm{bulk} = 4 \times \frac{8}{15} \epsilon_\textrm{CX} \approx 2 \epsilon_\textrm{CX}$. Similarly for syndrome qubits, but with an additional error associated with syndrome qubit readout, $q_\textrm{bulk} = 4 \times \frac{8}{15} \epsilon_\textrm{CX} + \epsilon_\textrm{M} \approx 2 \epsilon_\textrm{CX} + \epsilon_\textrm{M}$. To determine the dependence of the logical error probability on the code distance and local gate errors, we use a Monte Carlo simulation of errors ($p_\textrm{bulk}$ and $q_\textrm{bulk}$) and a local minimum weight perfect matching decoder \cite{Higgott2021}, exhibiting a threshold of 1.3\% (see main text Fig. 2).

Next, we consider the situation depicted in main text Fig. 1 where two surface code patches are being merged via teleported gates spanning the seam. 
As shown, stretching the parity check operators across the seam involves one teleported gate per (pink) data/syndrome qubit pair on the seam. Fig. \ref{circuit} shows how a teleported gate propagates bit and phase ($X$ and $Z$) errors occurring on a Bell pair.
Evidently, bit flip errors on the Bell pair propagate exclusively to the target qubit. 
Similarly, phase flip errors on the Bell pair propagate exclusively to the control qubit.
In total, the bit flip probability on the control qubit is $\frac{8}{15} \epsilon_\textrm{CX}$, whereas the bit flip probability on the target is $\frac{8}{15} \epsilon_\textrm{B} + 2 \times \frac{8}{15} \epsilon_\textrm{CX} + \epsilon_\textrm{M}$. 
Similarly, the phase flip probability on the control qubit is $\frac{8}{15} \epsilon_\textrm{B} + 2 \times \frac{8}{15} \epsilon_\textrm{CX} + \epsilon_\textrm{M}$, whereas the phase flip probability on the target is $\frac{8}{15} \epsilon_\textrm{CX}$.
Per code cycle, each (pink) qubit along the seam experiences three local CNOTs followed by one teleported gate.
As the seam shown in main text Fig. 1 is along the $X_L$ direction, it is most susceptible to logical bit flip errors $X_L$ arising from bit flips along the length of the seam, specifically on the pink qubits in code patch 2.
In contrast, although phase flip errors are also occurring with elevated probability along the length of the vertical seam (specifically on the pink qubits in code patch 1), they contribute little to logical phase errors, which correspond to horizontal strings.

For qubits on the seam (pink), our phenomenological weighted error model is as follows. A plaquette syndrome qubit on code patch (CP) 2 is the target of three local CNOTs ($3 \times \frac{8}{15} \epsilon_\textrm{CX}$), as well as the target of one teleported gate ($\frac{8}{15} \epsilon_\textrm{B} + 2 \times \frac{8}{15} \epsilon_\textrm{CX} + \epsilon_\textrm{M}$) and one final readout ($\epsilon_\textrm{M}$). The resulting bit flip probability for a plaquette syndrome qubit in code patch 2 is $q_{seam} = 3 \times \frac{8}{15} \epsilon_\textrm{CX} + (\frac{8}{15} \epsilon_\textrm{B} + 2 \times \frac{8}{15} \epsilon_\textrm{CX} + \epsilon_\textrm{M}) + \epsilon_\textrm{M} \approx 0.5 \epsilon_\textrm{B} + 2.5 \epsilon_\textrm{CX} + 2 \epsilon_\textrm{M}$. A data qubit on the seam of code patch 2 experiences identical bit flip probability to a plaquette syndrome qubit in the same patch, less one final readout, $p_{seam} = 3 \times \frac{8}{15} \epsilon_\textrm{CX} + (\frac{8}{15} \epsilon_\textrm{B} + 2 \times \frac{8}{15} \epsilon_\textrm{CX} + \epsilon_\textrm{M}) + \epsilon_\textrm{M} \approx 0.5 \epsilon_\textrm{B} + 2.5 \epsilon_\textrm{CX} + \epsilon_\textrm{M}$. Additionally, data qubits in code patch 2 experience phase flip errors at the same rate as bulk qubits. Correspondingly, similar considerations apply to data and star syndrome qubits in code patch 1, which experience elevated levels of phase flip errors. Table \ref{table1} in the main text summarizes the noise.

\subsection{Small Modules Error Model}
For comparison, in minimal-size module configuration, where each module contains a single data or syndrome qubit and just one communication qubit, teleported gates must be used for each two-qubit operation in all the check operators, leading to substantially worse performance. This configuration is illustrated in the cartoon in main text Fig. 2b.

Each code cycle, a given data qubit is the target of 2 teleported CNOTs and control of 2 teleported CNOTs, giving a bit flip probability per code cycle of $p_\textrm{small} = 2 \times (8\epsilon_\textrm{B}/15 + 2 \times 8\epsilon_\textrm{CX}/15 + \epsilon_\textrm{M}) + 2 \times (8\epsilon_\textrm{CX}/15) \approx \epsilon_\textrm{B} + 3 \epsilon_\textrm{CX} + 2 \epsilon_\textrm{M}$.

Similarly, consider a plaquette syndrome qubit, which is the target of four teleported CNOTs, then followed with additional readout error $\epsilon_\textrm{M}$. This plaquette has an error probability of $q_\textrm{small} = 4 \times (8\epsilon_\textrm{B}/15 + 2 \times 8\epsilon_\textrm{CX}/15 + \epsilon_\textrm{M}) + \epsilon_\textrm{M} \approx 2 \epsilon_\textrm{B} + 4 \epsilon_\textrm{CX} + 5 \epsilon_\textrm{M}$.
The result is the same for star operator syndrome qubits.

Also, since the threshold would occur at $p = 3 \%$ for $q = p$, expressing the threshold in terms of $\epsilon_\textrm{B}$ and $\epsilon_\textrm{CX}$ amounts to determining to what extent each of these contributes to $p$. In fact, taking for simplicity $\epsilon_\textrm{CX} = \epsilon_\textrm{M}$ and averaging the Small Modules expressions for $p$ and $q$ from Table \ref{sup_table1} gives a reasonable approximation for the threshold relationship $3 \% = 1.5 \epsilon_\textrm{B} + 7 \epsilon_\textrm{CX}$. An exact simulation with the Small Modules weights for $p$ and $q$ gives results quite close to this.

\begin{table}[h!]
\centering
\begin{tabular}{|c | c |c| c|} 
\hline
 & Bulk & Seam & Small Modules\\ 
 \hline
 $p$ & $2 \epsilon_\textrm{CX}$ & $0.5 \epsilon_\textrm{B}+2.5 \epsilon_\textrm{CX}+ \epsilon_\textrm{M}$ & $\epsilon_\textrm{B}+3 \epsilon_\textrm{CX}+2 \epsilon_\textrm{M}$ \\ 
 \hline
 $q$ & $2 \epsilon_\textrm{CX}+ \epsilon_\textrm{M}$ & $0.5 \epsilon_\textrm{B}+2.5 p_\textrm{CX}+2 \epsilon_\textrm{M}$ & $2\epsilon_\textrm{B}+4 \epsilon_\textrm{CX}+5 \epsilon_\textrm{M}$ \\
\hline
\end{tabular}
\caption{Phenomenological bit flip error probabilities per code cycle $p$ and $q$ on data and syndrome qubits. Entries describe how during a given code cycle, local operations and Bell pairs to the total phenomenological error probability. Phase flip error rates are identical. ``Bulk" and ``Seam" columns correspond to regions depicted in main text Fig. 1, and for comparison, the ``Small Modules" column shows the case where all gates in a surface code are done with teleported gates. }
\label{sup_table1}
\end{table}

\subsection{Cavity-Enhanced Optical Pumping}

We perform full simulations of Rubidium-87 with both D1 and D2 lines. We consider a degenerate cavity with a resonance on the D1 transition $F=1 \leftrightarrow F'=1 $ and a second resonance on the D2 $F=1 \leftrightarrow F'=0 $ transition. With the quantization axis along the cavity axis, the polarization modes of the cavity are $\sigma_+$ and $\sigma_-$ polarization. This field configuration and cavity resonances are used for both atom-photon entanglement generation and cavity-enhanced optical pumping.
\begin{figure*}[h]
\includegraphics[width=.75\textwidth]{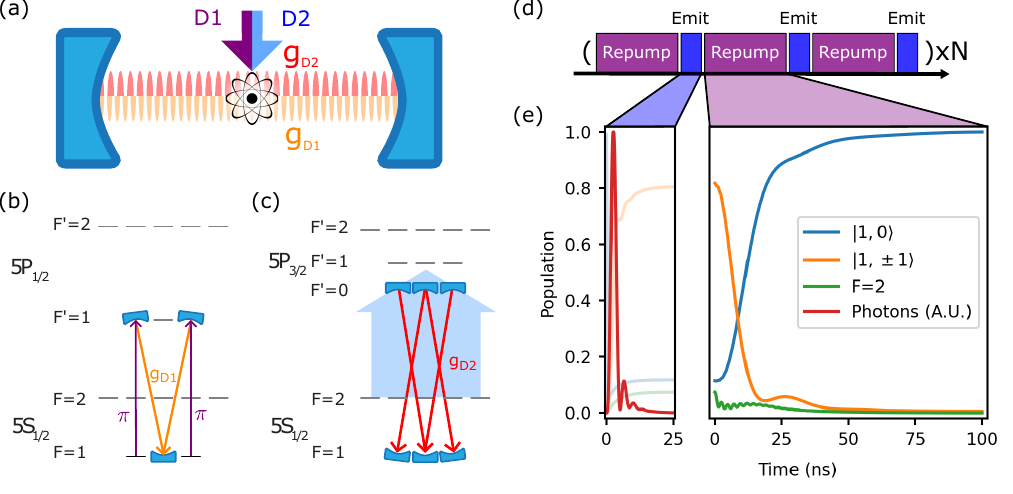}
\caption{(a) The atom is placed inside an optical cavity with resonances on the D1 (yellow) and D2 (red) transitions with a coupling strength $g_{D1}$ and $g_{D2}$, respectively. Drives on the D1 (purple) and D2 (blue) transitions are sent from the side of to perform optical pumping on the atom. (b) A level diagram showing rapid pumping into the clock state through the cavity on the D1 transition. (c) A level diagram of the D2 drive pumping out of the F=2 manifold back into the F=1 manifold. (d) a sequence in time showing consecutive repump, and photon emission. (e) Population dynamics are shown as a function of time. Photon emission is performed, if entanglement is not successfully generated, this leaves the atom in a mixture of states. The following plot shows rapid cavity enhanced repumping back into to $|1,0\rangle$ state. }
\label{enhancedrepump}
\end{figure*}
Cavity-enhanced optical pumping is performed by applying a linearly polarized drive on the D1 transition $F=1 \leftrightarrow F'=1 $ transition while simultaneously applying a drive on the D2 $F=1 \leftrightarrow F'=0 $ transition. The drive on the D1 pumps atoms into the $|F=0, m_F = 0\rangle $ due to selection rules forbidding $m_{F'} = 0\leftrightarrow m_F = 0 $ transitions. The drive on the D2 line pumps atoms from the $F=2$ manifold into the $F=1$ manifold. Due to the cavity coupling, both drives perform pumping at enhanced rates. We verify the effectiveness of the cavity-enhanced optical pumping by preparing an atom in an equal mixture of all ground states and observing the rate of pumping into the $|F=0, m_F = 0\rangle $ state. 
The finite hyperfine splitting in the $5P_{1/2}$ results in the D1 drive off resonantly scattering from the $|F'=2, m_F = 0\rangle $ state. This creates a tradeoff between pumping speed and fidelity. Pumping is optimized by setting a constant timescale of $100$ ns and then optimize drive parameters for the pumping fidelity. Optimal parameters are found with a D2 rabi frequency of $330$ MHz on the D2 line and  $50$ MHz on the D1 line. After $100$ ns the atom is pumped into the $|F=0, m_F = 0\rangle $ with 99.4$\%$ fidelity.

\subsection{Rate Comparison}
In main text Fig. 3, the remote-Bell pair rate for all approaches is the entanglement attempt rate $R_{\textrm{attempt}}$ multiplied by atom-atom entanglement generation probability per attempt $P_\textrm{aa}$. The attempt rate increases with qubit number through a combination of temporal and spatial multiplexing, until it is limited, when applicable, by the number of optical modes. The success probabilities and attempt rates for each approach are summarized in Table \ref{sup_table2}.

\begin{table}[h!]
\centering
\begin{tabular}{|c | c | c|} 
\hline
 &  $P_\textrm{aa}$ & $R_{\textrm{attempt}}(N)$  \\ 
 \hline
 single cavity & 0.1 & Min[$N/(16 \: \mu$s), 10 $\mu$s$^{-1}$] \\ 
 \hline
  30 micro-cavities & 0.24 & Min[$N/(0.3 \: \mu$s), 286 $\mu$s$^{-1}$] \\
\hline
free-space & 0.0035 & $N/(16 \: \mu$s) \\
\hline
\end{tabular}
\caption{Entanglement generation success probability and attempt rate for each approach described in the main text. Multiplying $P_\textrm{aa}$ by $R_{\textrm{attempt}}(N)$ yields the curves shown in main text Fig. 3. }
\label{sup_table2}
\end{table}

\end{document}